\documentclass[12pt]{article}
\usepackage{graphicx}
\begin{document}

\title{Tagged--photon events in polarized DIS process}

\author{G.I. Gakh, M.I. Konchatnij, and N.P. Merenkov}

\date{}
\maketitle
\begin{center}
{\small {\it  {NSC "Kharkov Institute of Physics and Technology"  \\ }}}
{\small {\it {61108, Akademicheskaya 1, Kharkov, Ukraine}}}
\end{center}

\begin{abstract}
Deep--inelastic events for the scattering of the longitudinally
polarized electron by polarized proton with tagged collinear
photon radiated from initial--state electron are considered. The
corresponding cross--section is derived in the Born approximation.
The model--independent radiative corrections to the Born
cross--section are also calculated. Obtained result is applied to
the case of elastic scattering.
\end{abstract}

\section{Introduction}

\hspace{0.7cm}
The idea to use radiative events in lepton--hadron interaction to expand
the experimental possibilities for studies of different topics in the
high--energy physics has become quite attractive in the last years.

Photon radiation from the initial $e^+e^-$--state, in the events
with missing energy, has been successfully used at LEP for the
measurement of the number of light neutrinos and for search the
new physics signals \cite {Acc}. The possibility to undertake the
bottonium spectroscopy studies at $B$--factories by using emission
of a hard photon from the electron or the positron has been
considered in Ref. \cite{BEIS}. The important physical problem of
the the total hadronic cross--section scanning in the
electron--positron annihilation process at low and intermediate
energies by means of the initial--state radiative events has been
discussed widely in Ref. \cite{scanning}.

The initial--state collinear radiation is very important in
certain regions of the deep--inelastic scattering (DIS) at $HERA$
kinematic domain. It leads to reduction of the projectile electron
energy and therefore to a shift of the effective Bjorken variables
in the hard scattering process as compared to those determined
from the actual measurement of the scattered electron alone. That
is why the radiative events in the DIS process
\begin{equation}\label{1,radiative process}
e^-(k_1) +p(p_1) \rightarrow e^-(k_2) + \gamma(k) + X(p_x)
\end{equation}
have to be carefully taken into account \cite{MT}.

Besides, the measurement of the energy of the photon emitted very
close to the incident electron beam direction \cite{KPS,JJP}
permits to overlap the kinematical region of photoproduction
$(Q^2=-(k_1-k_2)^2 \simeq 0)$ and DIS region with small trasferred
momenta $(Q^2$ about a few $GeV^2$) within the high--energy $HERA$
experiments. These radiative events may be used also to determine
independently the proton structure functions $F_1$ and $F_2$ in a
single run without lowering the beam energy \cite{KPS,FG}. The
high--precision calculation (taking into account radiative
corrections (RC)) of the corresponding cross--section has been
performed in Ref. \cite{AAKM}.

In the present paper we investigate the deep--inelastic events for
the radiative process (1) with longitudinally polarized electron
beam and polarized proton as a target. We suggest that, as in Ref.
\cite{AAKM}, the hard photon is emitted very close to the
direction of the incoming electron beam $(\theta_{\gamma} = {\bf
\widehat{p_1k_1}} \leq\theta_0, \ \theta_0\ll 1),$ and the photon
detector (PD) measures the energy of all photons inside the narrow
cone with the opening angle $2\theta$ around the electron beam.
Simultaneously the scattered--electron 3--momentum is fixed.

We consider the cases of the longitudinal (along the electron beam
direction) and perpendicular (in the plane $({\bf k_1,k_2})$)
polarizations of the proton. In Section 2 we derive the
corresponding cross--sections in the Born approximation and in
Section 3 we calculate the different contributions into RC to the
Born cross--section. The total radiative correction for different
(exclusive and calorimeter) experimental conditions for the
scattered--electron measurement is given in Section 4.  Our
results can be applied to describe the cross--section of the
process (1) with target proton at rest as well as with colliding
electron--proton beams. In Section 5 we apply the obtained in
Section 4 results to describe the quasi--elastic scattering using
the connection between the spin--dependent proton structure
functions and the proton electromagnetic form factors in this
limiting case.

\section{Born approximation}

\hspace{0.7cm}

The spin--independent part of the DIS cross--section with
considered here experimental set--up has been investigated
recently in details \cite{AAKM}. Now we consider the
spin--dependent part of the corresponding cross--section that is
described by means of the proton structure functions $g_1$ and
$g_2.$ As the opening angle of the forward PD is very small, and
we consider only the cross--section where the tagged photon is
integrated over the solid angle covered by PD, we can apply the
quasi--real electron method \cite{BFK} and parametrize these
radiative events using the standard Bjorken variables
\begin{equation}\label{2,standard variables}
x=\frac{Q^2}{2p_1(k_1-k_2)}\ , \ \ y=\frac{2p_1(k_1-k_2)}{V}\ , \
\ V=2p_1k_1 \ ,
\end{equation} and the energy fraction of the electron after the
initial--state radiation of a collinear photon
\begin{equation}\label{3}
z=\frac{2p_1(k_1-k)}{V} = \frac{\varepsilon-\omega}{\varepsilon}\ ,
\end{equation}
where $\varepsilon$ is the initial--electron energy and $\omega$
is the energy deposited in PD.

An alternative set of the kinematic variables, that is specially
adapted to the case of the collinear--photon radiation, is given
by the shifted Bjorken variables \cite{KPS,KMF}
\begin{equation}\label{4,shifted variables} \widehat Q^2 =
-(k_1-k_2-k)^2\ , \ \ \hat x = \frac{\widehat
Q^2}{2p_1(k_1-k_2-k)}\ , \ \ \hat y
=\frac{2p_1(k_1-k_2-k)}{2p_1(k_1-k)} \ .
\end{equation}
The relation between the shifted and standard Bjorken variables reads
\begin{equation}\label{5,relations}
\widehat Q^2 = zQ^2\ , \ \ \hat x = \frac{xyz}{z+y-1}\ , \ \ \hat y =
\frac{z+y-1}{z} \ .
\end{equation}
At fixed values of $x$ and $y$ the lower limit of $z$ can be derived from
constraint on the shifted variable $\hat x$
$$\hat x < 1 \ \ \ \rightarrow \ \  z  >  \frac{1-y}{1-xy} \ . $$

In the framework of the Born approximation we use the following
determination of the DIS cross--section in the radiative process
(1) in terms of the contraction of the leptonic and hadronic
tensors (further we will interest with the spin--dependent part of
the cross--section only)
\begin{equation}\label{6,definition Born}
\frac{d\sigma}{\hat y d\hat x d\hat y} = \frac{4\pi\alpha^2(\widehat
Q^2)}{\widehat Q^4}zL_{\mu\nu}^BH_{\mu\nu}\ ,
\end{equation}
where $\alpha(\widehat Q^2)$ is the running electromagnetic
coupling constant, that takes into account the effects of the vacuum
polarization, and the Born leptonic current tensor in considered case
reads \cite{KS} \begin{equation}\label{7,Born leptonic tensor}
L_{\mu\nu}^B =
\frac{\alpha}{4\pi^2}\int_{\Omega}\frac{d^3k}{\omega}2i\varepsilon_{\mu\nu\lambda\rho}
q_{\lambda}\bigl(k_{1\rho} R_t + k_{2\rho}R_s\bigr)\ , \ q=k_1-k_2-k\ ,
\end{equation}
where $\Omega$ covers the solid angle of $PD.$

For the case of the initial--state collinear radiation, which we
cosider in this paper, quantities $R_t$ and $R_s$ can be written
as follows
\begin{equation}\label{8}
R_t=-\frac{1}{(1-z)t}-\frac{2m^2}{t^2}\ , \ \ R_s =
-\frac{z}{(1-z)t}+\frac{2m^2(1-z)}{t^2}\ , \ \ t=-2kk_1\ , \ \ q=zk _1-k_2
\ .\end{equation}

The trivial angular integration of the Born leptonic tensor
gives in accordance with the quasi--real electron
approximation \cite{BFK} \begin{equation}\label{9} L_{\mu\nu}^B =
\frac{\alpha}{2\pi}P(z,L_0)dzi\varepsilon_{\mu\nu\lambda\rho}
q_{\lambda}k_{1\rho}\ , \ \  L_0 = \ln\frac{\varepsilon^2\theta_0^2}{m^2}\
, \end{equation}
$$P(z,L_0) = \frac{1+z^2}{1-z}L_0 -\frac{2(1-z+z^2)}{1-z}\ , $$
where $m$ is the electron mass.

We write the spin--dependent part of the hadronic tensor, on the
right side of Eq.~(6), in the following form
\begin{equation}\label{10,hadronic tensor} H_{\mu\nu} =
-iM\frac{\varepsilon_{\mu\nu\lambda\rho}q_{\lambda}}{2p_1q}
\bigl[(g_1+g_2)S_{\rho}-g_2\frac{Sq}{p_1q}p_{1\rho}\bigr]\ ,
\end{equation}
where $M$ is the proton mass and $S$ is the 4--vector of the proton
polarization. When writing the expressions (7) and (10), we suppose that
the polarization degree of both the electron and the proton equals to $1.$

Our normalization is such that the proton structure functions $g_1$ and
$g_2$ are dimensionless and in the limiting case of the elastic scattering
$(\hat x\rightarrow 1)$ they are expressed in terms of the proton electric
$(G_E)$ and magnetic $(G_M)$ form factors as follows
\begin{equation}\label{11}
g_1(\hat x,\widehat Q^2)\rightarrow  \delta(1-\hat x)\bigl[G_MG_E+
\frac{\lambda}{1+\lambda}(G_M-G_E)G_M\bigr]\ , \ \lambda=\frac{\widehat
Q^2}{4M^2} \ ,
\end{equation}
$$ g_2(\hat x,\widehat Q^2)\rightarrow - \delta(1-\hat x)
\frac{\lambda}{1+\lambda}(G_M-G_E)G_M\ , \ \ G_{M,E} = G_{M,E}(\widehat
Q^2) \ . $$

In our problem it is convenient to parametrize the 4--vector of the
proton polarization in terms of the 4--momenta of the reaction
under study \cite{AS} \begin{equation}\label{12,prametrization of S}
S_{\mu}^{\parallel} = \frac{2M^2k_{1\mu}-Vp_{1\mu}}{MV}\ , \ \
S_{\mu}^{\bot} =
\frac{up_{1\mu}+Vk_{2\mu}-[2u\tau+V(1-y)]k_{1\mu}}{\sqrt{-uV^2(1-y)-u^2M^2}}
\ ,
\end{equation}
where $u=-Q^2, \ \tau = M^2/V$ and we neglect the electron mass
here. The 4--vector of the longitudinal proton polarization has
components
\begin{equation}\label{13, components long}
S_{\mu}^{\parallel} = (0, {\bf n_1})\ , \ \ S_{\mu}^{\parallel} =
\Bigl(-\frac{|{\bf p_1}|}{M}, \frac{{\bf n_1}E_1}{M}\Bigr)
\end{equation}
for the target at rest and colliding beams, respectively. Here
$E_1({\bf p_1})$ is the proton energy (3--momentum) and ${\bf
n_1}$ is the unit vector along the initial--electron 3--momentum
direction. The 4--vector of the perpendicular proton polarization
$S_{\mu}^{\bot}$ is the same for both these cases
\begin{equation}\label{14,components perp}
S_{\mu}^{\bot} = \Bigl(0, \ \frac{{\bf n_2-n_1(n_1n_2)}}{\sqrt{1-{\bf
(n_1n_2)^2}}}\Bigr)\ ,
\end{equation}
where ${\bf n_2}$ is the unit vector along the scattered--electron
3--momentum direction. It is easy to verify that
$S^{\parallel}S^{\bot} =0.$

 \begin{figure}[h]
\includegraphics[width=0.32\textwidth]{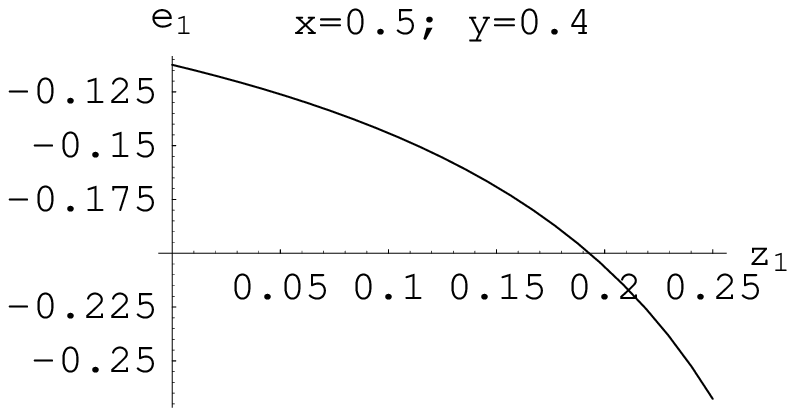}
 \hfill
\includegraphics[width=0.32\textwidth]{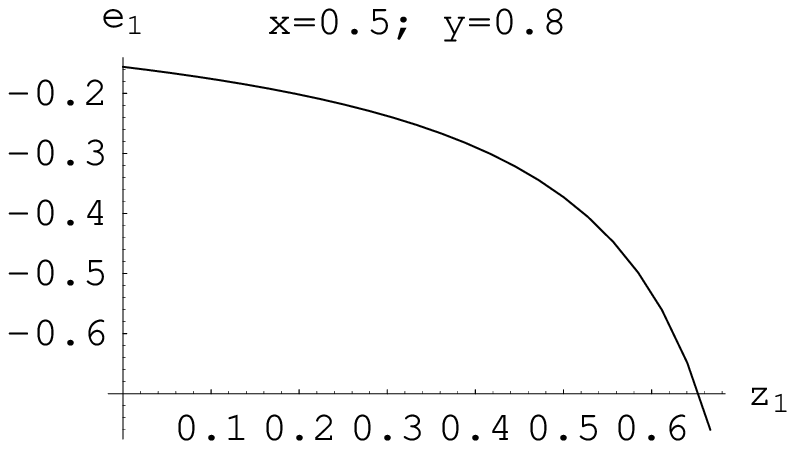}
 \hfill
\includegraphics[width=0.32\textwidth]{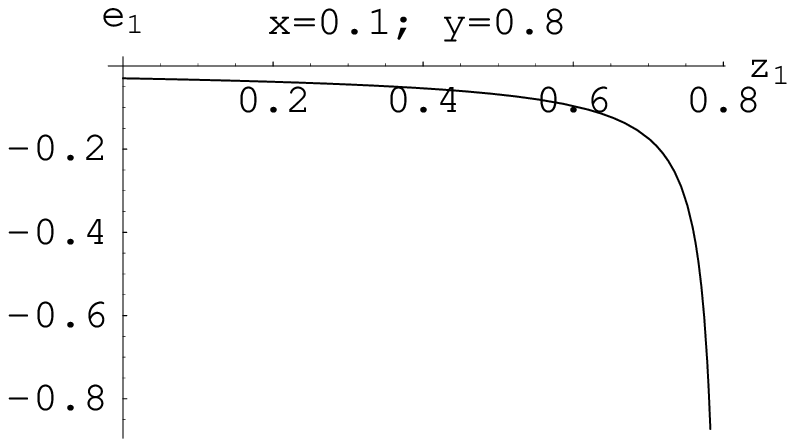}
\hfill
\includegraphics[width=0.32\textwidth]{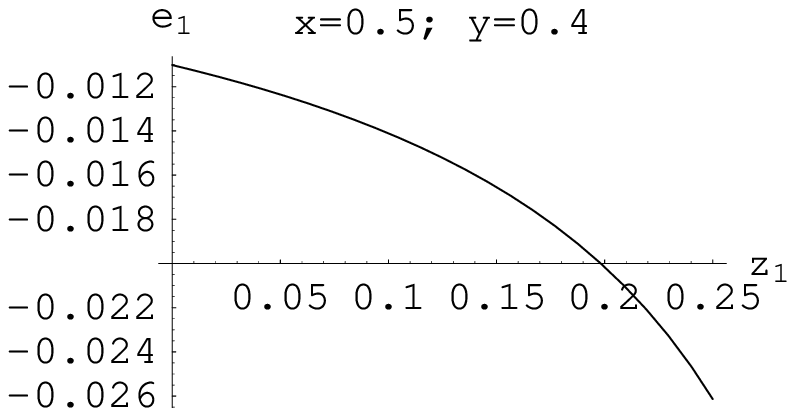}
\hfill
\includegraphics[width=0.32\textwidth]{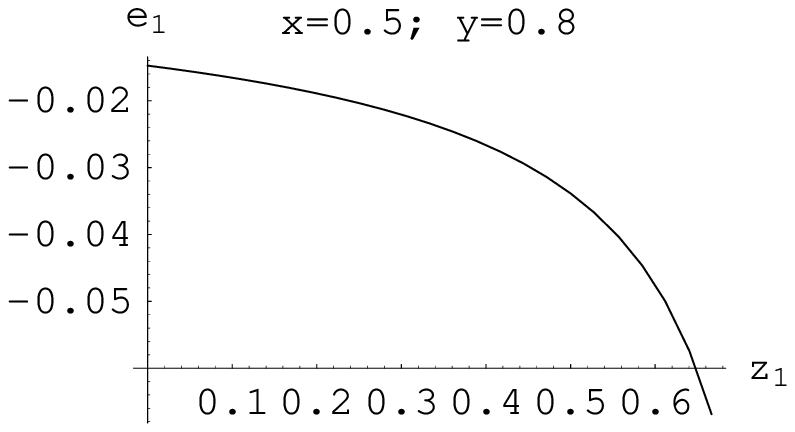}
\hfill
\includegraphics[width=0.32\textwidth]{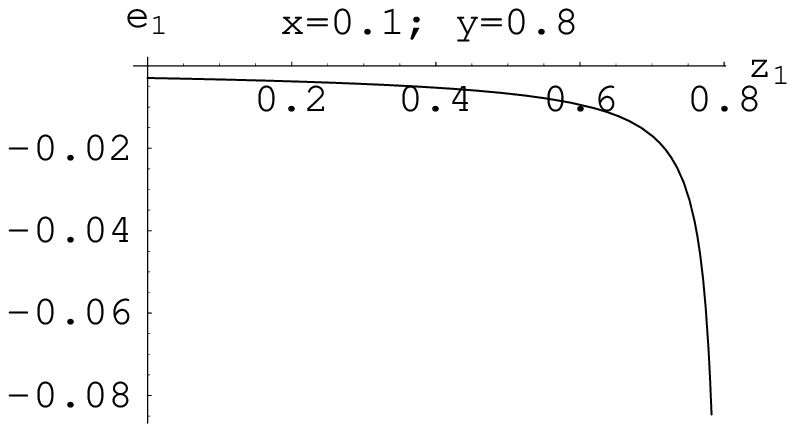}
\\
 \parbox[t]{1\textwidth}{\caption{The dependence of the quantity $e_1$ on the energy
 fraction of the tagged photon $z_1=1-z$ for different values of $x,\ y$ and V. The upper
 set corresponds to $V=10$ $GeV^2$ and the lower one to $V=100$ $GeV^2.$
 The maximum value of $z_1$ is $y(1-x)/(1-xy).$}\label{fig1}}

\end{figure}

Using the definitions of the DIS cross--section (6), leptonic and
hadronic tensors (9), (10) and parametrization of the proton
polarization (12), after simple calculations, we derive the
spin--dependent part of the cross--section of the process (1),
with tagged collinear photon radiated from initial state, in the
following form
\begin{equation}\label{15,born cross section}
\frac{d\sigma_{\parallel,\bot}^B}{\hat y d\hat{x}d\hat{y}dz} =
\frac{\alpha}{2\pi} P(z,L_0)\Sigma_{\parallel,\bot}(\hat x,\hat
y,\widehat Q^2)\ , \end{equation} \begin{equation}\label{16}
\Sigma_{\parallel} = \frac{4\pi\alpha^2(\widehat Q^2)}{\widehat
V\hat y} \bigl(\hat\tau-\frac{2-\hat y}{2\hat x\hat y}\bigr)
g_1(\hat x, \widehat Q^2)[1+ e_1\hat R(\hat x,\widehat
Q^2)] \ ,
\end{equation}
\begin{equation}\label{17}
\Sigma_{\bot} = -\frac{4\pi\alpha^2(\widehat Q^2)}{\widehat V\hat y}
\sqrt{\frac{M^2}{\widehat Q^2}(1-\hat y-\hat x\hat y\hat\tau)}
g_1(\hat x, \widehat Q^2)[1+
e_2\hat R(\hat x,\widehat Q^2)]\ ,
\end{equation}
where
$$e_1 = \frac{4\hat\tau\hat x}{2\hat x\hat
y\hat\tau+\hat y-2} \ , \ e_2 = \frac{2}{\hat y} \ ,
\ \widehat R = \frac{g_2(\hat x,\widehat Q^2)}{g_1(\hat x,\widehat Q^2)}\
, \ \hat\tau =\frac{M^2}{\widehat V}\ , \ \widehat V = zV\ . $$

It is useful to remind that unpolarized $DIS$ cross--section is
proportional to $\sigma_T(1+eR)$, where
$R=\sigma_L/\sigma_T$ and for events with the tagged collinear
photon [5] $$e = \frac{2(1-\hat y)}{1+(1-\hat y)^2} \ . $$

\vspace{0.1cm}
 \begin{figure}[h]
\includegraphics[width=0.32\textwidth]{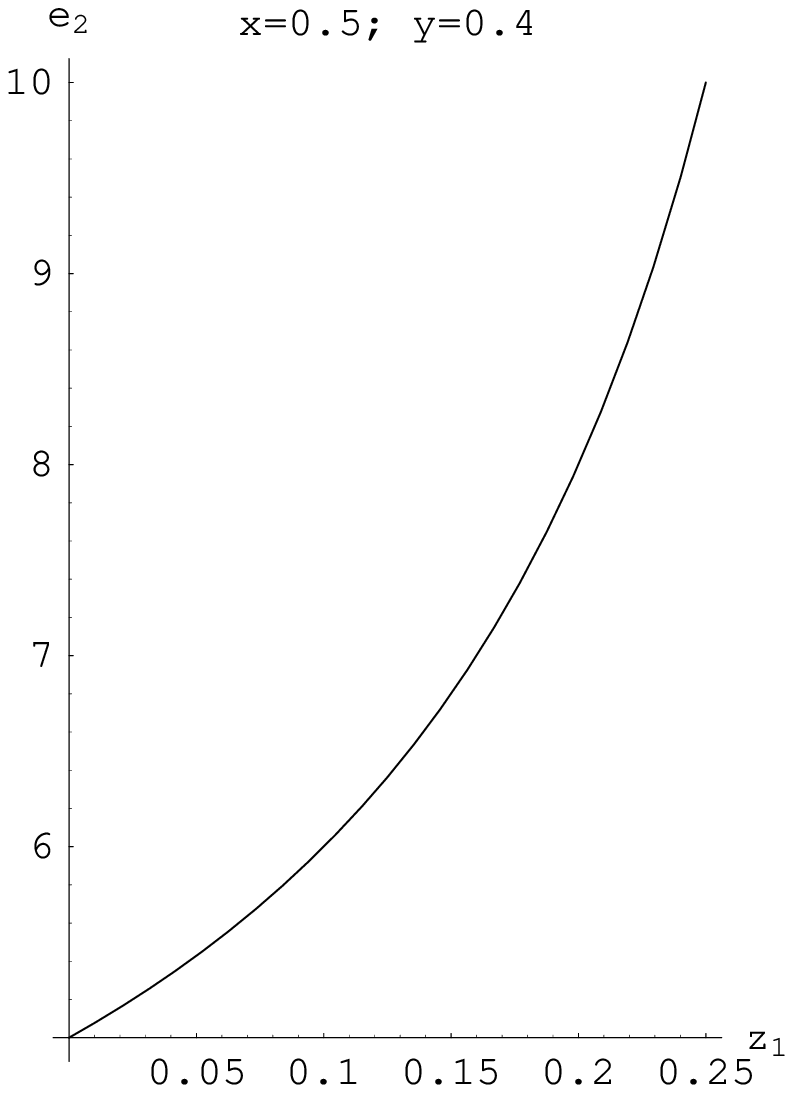}
 \hfill
\includegraphics[width=0.32\textwidth]{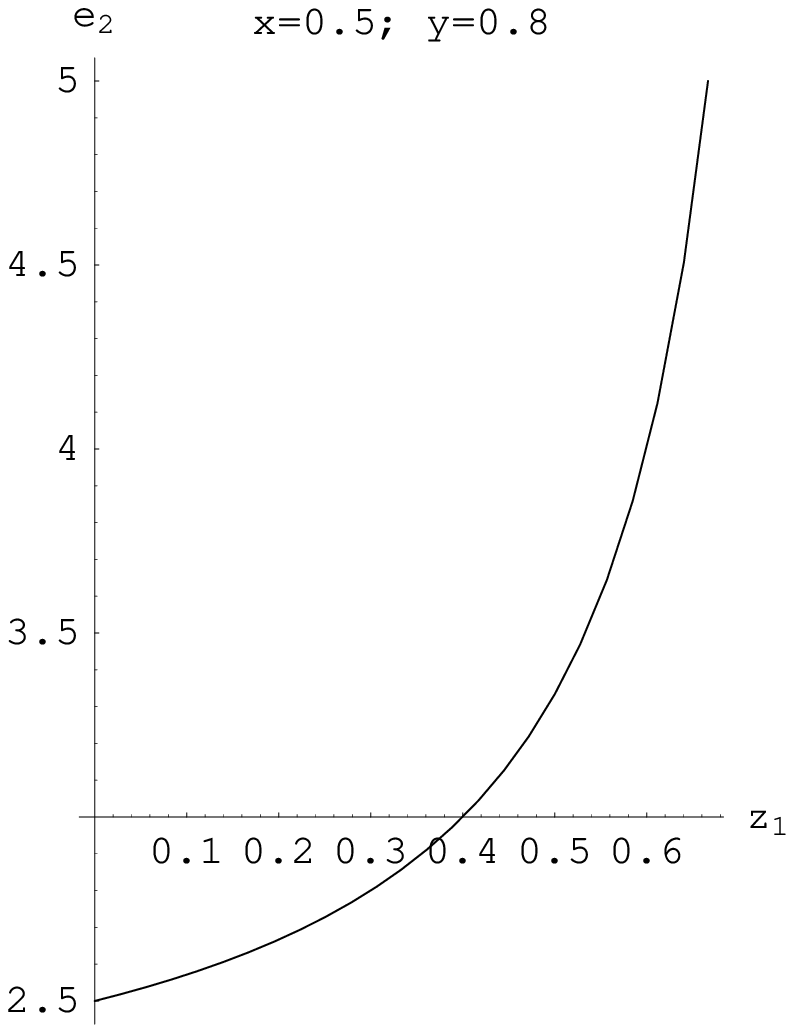}
 \hfill
\includegraphics[width=0.32\textwidth]{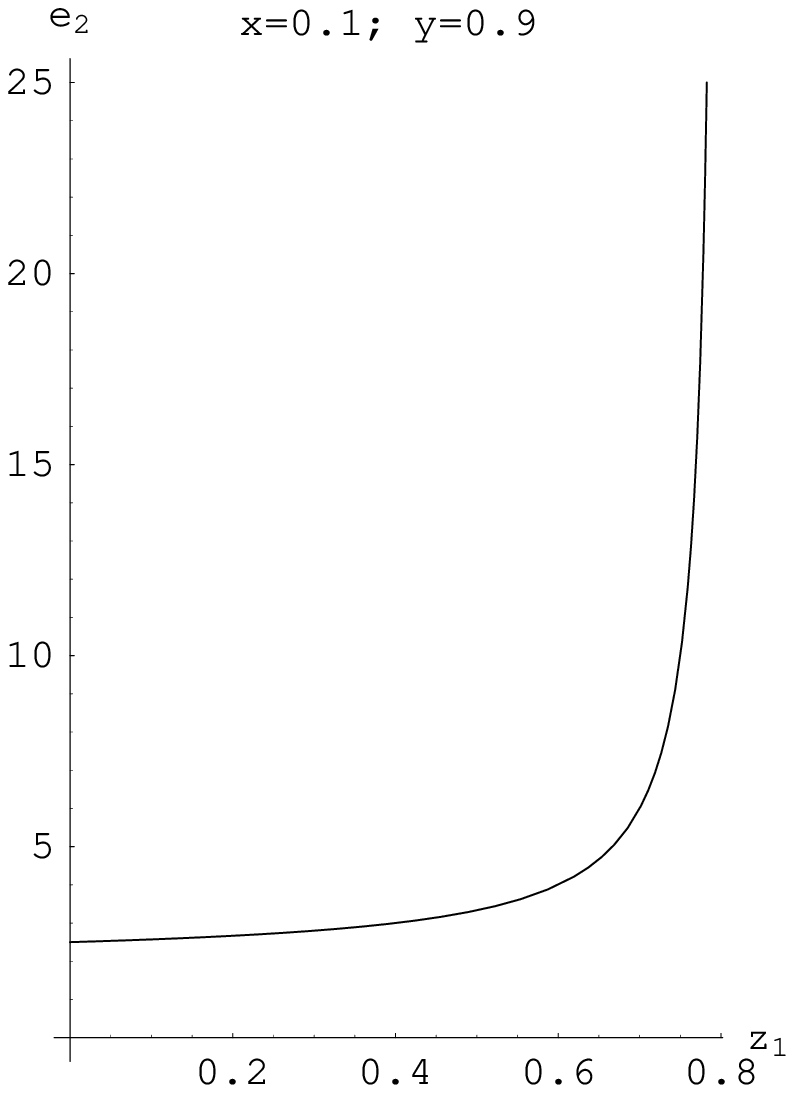}
\\
 \parbox[t]{1\textwidth}{\caption{The quantity $e_2$ at different values
 of $x$ and $y$ as the function of $z_1=1-z$.}\label{fig2}}

\end{figure}

Because the quantities $e_1$ and $e_2$ depend strongly on $z$, the
determination of the proton structure functions $g_1$ and $g_2$ is
possible by measurement of $z$--dependence of the cross--section
(15) in a single run without lowering the electron beam energy.
The quantity $e_1$ is proportional to $\tau $ and, therefore, is
very small at the $HERA$ conditions. Thus, the separation of the
$g_1$ and $g_2$ in the $DIS$ process with the longitudinally
polarized proton is possible in experiments with the target at
rest and low values of V (up to 20 $GeV^2$). At $HERA$ the
cross--section of this process can be used to measure the
structure function $g_1$ only. This can be seen from Fig.~1.
Contrary, the Fig.~2 shows that the experiments with the tagged
photon and perpendicular polarization of the proton can be used to
measure both $g_1$ and $g_2$ in the wide interval of the energies
(provided quantity $Q^2$ is not large).

\section{Radiative corrections}

\hspace{0.7cm}

We will restrict ourselves to the model--independent QED radiative
corrections related to the radiation of the real and virtual
photons by leptons. The remaining sources of RC in the same order
of the perturbation theory, such as the virtual corrections with
double photon exchange mechanism and bremsstrahlung off the proton
and partons, are more involved and model dependent. They are not
considered here. Our approach to the calculation of RC is based on
the account of all essential Feyman diagrams that describe the
observed cross--section in framework of the used approximation. To
get rid of cumbersome expressions we will retain in RC the terms
that accompanied at least by one power of large logarithms.  In
the considered case three different types of such logarithms
appear \begin{equation}\label{18, logarithms} L_0\ , \ \ L_Q =
\ln\frac{Q^2}{m^2}\ , \ \ L_{\theta} = \ln\frac{\theta_0^2}{4}\ .
\end{equation}
Besides, in chosen approximation we neglect the terms of the order of
$\theta_0^2$, $m^2/\varepsilon^2\theta_0^2$ and $m^2/Q^2$ in the
cross--section.

The total RC to the cross--section (15) includes the contributions due to
the virtual and soft photon emission as well as hard photon radiation. We
begin with the calculation of the virtual and soft corrections.

\subsection{Virtual and soft corrections}

\hspace{0.7cm}

To calculate the contribution from the virtual-- and soft--photon
emission corrections, we start from the expression for the
one--loop corrected Compton tensor with a heavy photon for
longitudinally polarized electron \cite{AAK}. For considered here
hard collinear initial--state radiation this Compton tensor can be
written as
\begin{equation}\label{19, Virtual tensor}
L_{\mu\nu}^V = \frac{\alpha}{2\pi}\rho L_{\mu\nu}^B
+\frac{\alpha^2}{4\pi^3}\int_{\Omega}
i\varepsilon_{\mu\nu\lambda\rho}q_{\lambda}k_{1\rho}\frac{d^3k}{\omega}
\Bigl[\frac{T}{-t}+\frac{4m^2(1-z+z^2)}{t^2}L_Q\ln z\Bigr] \ ,
\end{equation}
$$
T=\frac{1+z^2}{1-z}[2\ln{z}(l_t-\ln(1-z)-L_Q)-2F(z)]+\frac{1+2z-z^2}{2(1-z)}
\ , \ \  F(z) = \int\limits_1^{1/z}\frac{dx}{x} \ln|1-x| \ , $$
$$l_t=\ln\frac{-t}{m^2}\ ,\ \ \rho =
4(L_Q-1)\ln\frac{\delta}{m}-L_Q^2 +3L_Q+3\ln z
+\frac{\pi^2}{3}-\frac{9}{2}\ , $$ where $\delta$ is the fictitious
photon mass, and tensor $L^B_{\mu\nu}$ is defined by Eq.~(9).

To eliminate the photon mass we have to add the contribution due
to additional soft--photon emission with the energy less than
$\Delta\varepsilon\ , \ \Delta\ll 1.$ This contribution has been
found in Ref. \cite{HV} and the corresponding procedure of the
photon--mass elimination has been described in Ref. \cite{KMF1}.
The result reads
\begin{equation}\label{20}
L_{\mu\nu}^{V+S} = L_{\mu\nu}^V(\rho\rightarrow\tilde\rho)\ ,
\end{equation}
$$\tilde\rho =2(L_Q-1)\ln\frac{\Delta^2}{Y}+3L_Q+3\ln{z}-\ln^2{Y} -\frac
{\pi^2}{3}-\frac{9}{2}+2Li_{2}(\cos^2\frac{\theta}{2})\ , \ \ Y=\frac
{\varepsilon_2}{\varepsilon}\ , $$
where $\varepsilon_2$ is the scattered--electron energy and $\theta$ is
the electron scattering angle ($\theta = {\bf \widehat{k_1k_2}}$).

The angular integration with respect to the hard tagged photon over the
solid angle of PD gives (in the framework of used accuracy)
\begin{equation}\label{21}
\L_{\mu\nu}^{V+S} = \Bigl(\frac{\alpha}{2\pi}\Bigr)^2\bigl[\tilde\rho
P(z,L_0)+
G\bigr]dzi\varepsilon_{\mu\nu\lambda\rho}q_{\lambda}k_{1\rho} \ ,
\end{equation}
$$G=\Bigl\{\frac{1+z^2}{1-z}[\ln{z}(L_0-2L_Q)-2F(z)]+\frac{1+2z-z^2}{2(1-z)}
\Bigr\}L_0 + \frac{4(1-z+z^2)}{1-z}L_Q\ln{z} \ .$$

Using the right side of Eq.~(21) insdead of $L_{\mu\nu}^B$ on the right
side of Eq.~(6), we derive the contribution of the virtual and soft
corrections to the Born cross--section (15) in the following form
\begin{equation}\label{22, V+S}
\frac{d\sigma^{V+S}_{\parallel,\bot}}{\hat y d\hat yd\hat x
dz}=\Bigl(\frac{\alpha}{2\pi}\Bigr)^2[\tilde\rho P(z,L_0)+G]
\Sigma_{\parallel,\bot}(\hat x,\hat y, \widehat Q^2) \ ,
\end{equation}
where $\Sigma_{\parallel,\bot}(\hat x,\hat y, \widehat Q^2)$ are
defined by Eqs.~(16), (17).

\subsection{Double hard bremsstrahlung}

\hspace{0.7cm}

Let us consider the emission of an additional hard photon with
4--momentum $\tilde k$ and the energy more than $\Delta\varepsilon$. To
calculate the contribution from the real hard bremsstrahlung, which in our
case corresponds to double hard photon emission, with at least one photon
seen in the forward PD, we specify three specific kinematical domains:

$i)$ both hard photons hit the forward PD, i.e. both are emitted within a
narrow cone around the electron beam $({\bf \widehat{kk_1}, \widehat{\tilde
kk_1}}\leq\theta_0);$

$ii)$ one hard photon is tagged by PD, while the other one is collinear to
the outgoing electron momentum $({\bf \widehat{\tilde kk_2}}\leq\theta'_0\
, \ \theta'_0\ll 1);$

$iii)$ the additional photon is emitted at large angles (i.e. outside the
both defined narrow cones) with respect to both incoming and outgoing
electron momenta.

The contributions of the regions $i)$ and $ii)$ contain terms
quadratic in the large logarithms $L_0, \ L_Q$, whereas region
$iii)$ contains terms of the order of $L_0L_{\theta}$, which can
give numerically even larger contribution if
$2\theta_0>\varepsilon\theta_0/m.$

We denote the third kinematic domain as a semi--collinear one.
Beyond the leading logarithmic accuracy, the calculation may be performed
using the results of paper \cite{KM} for leptonic current tensor with
longitudinally polarized electron for collinear as well as semi--collinear
regions.

The contribution from the kinematic region $i)$, when both hard photons
hit PD and every one has the energy more than $\Delta\varepsilon,$ can be
written as follows
\begin{equation}\label{23, two photons hit PD}
\frac{d\sigma^{i)}_{\parallel,\bot}}{\hat y d\hat yd\hat xdz} =
\Bigl(\frac{\alpha}{2\pi}\Bigr)^2L_0\Bigl\{\bigl[\frac{1}{2}P^{(2)}_{\theta}(z)
+\frac{1+z^2}{1-z}\bigl(\ln{z}-\frac{3}{2}-2\ln\Delta\bigr)\bigr]L_0 +
\end{equation}
$$7(1-z)-2(1-z)\ln{z}+\frac{3+z^2}{2(1-z)}\ln^2z-2\frac{3-2z+3z^2}{1-z}\ln
\frac{1-z}{\Delta}\Bigr\}\Sigma_{\parallel,\bot}(\hat x,\hat
y,\widehat Q^2) \ .  $$ The double--logarithmic terms on the right
side of Eq.~(23) are the same for the polarized and unpolarized
cases, whereas one--logarithmic terms are different. In Eq.~(23)
we use the notation $P^{(2)}_{\theta}(z)$ for the $\Theta$--part
of the second--order electron structure function $D(z,L)$
\cite{jad}
$$D(z,L)=\delta(1-z)+\frac{\alpha}{2\pi}P^{(1)}(z)L+\frac{1}{2}\Bigl(
 \frac{\alpha}{2\pi}\Bigr)^2P^{(2)}(z)L^2 + ... \ ,$$
$$P^{(i)}(z)=P^{(i)}_{\theta}(z)\Theta(1-z-\Delta)+\delta(1-z)
P^{(i)}_{\delta}\ , \ \ \Delta\rightarrow 0\ , $$
$$P^{(1)}_{\theta}(z)=\frac{1+z^2}{1-z}\ , \ \ P^{(1)}_{\delta} =
\frac{3}{2}+2\ln\Delta\ , $$
\begin{equation}\label{24, D--function}
P^{(2)}_{\theta}(z)=2\Bigl[\frac{1+z^2}{1-z}\Bigl(2\ln(1-z)-\ln{z}+
\frac{3}{2}\Bigr)+\frac{1}{2}(1+z)\ln{z}-1+z\Bigr] \ .
\end{equation}

To calculate the contribution of the kinematical region $ii)$ we can use
the quasi--real electron method to describe the radiation of both collinear
photons. This contribution to the observed cross--section depends on the
event selection, in other words, on the method of measurement of the
scattered electron.

For exclusive event selection, when only the scattered electron is
detected, while the photon, that is emitted almost collineary (i.e. within
the opening angle $2\theta'_0$ around the momentum of the scattered
electron), goes unnoticed or is not taken into account in the determination
of the kinematical variables, we have in accordance with Ref. \cite{BFK}
\begin{equation}\label{25, incl.event selection}
\frac{d\sigma^{ii),excl}_{\parallel,\bot}}{\hat y d\hat yd\hat xdz} =
\frac{\alpha^2}{4\pi^2}P(z,L_0)\int\limits_{\Delta/Y}^{y_{1max}}\frac{dy_1}
{1+y_1}\Bigl[\frac{1+(1+y_1)^2}{y_1}(\widetilde
L-1)+y_1\Bigr]\Sigma_{\parallel,\bot}(x_s,y_s,Q_s^2)\ ,
\end{equation}
where $y_1$ is the energy fraction of the photon, radiated along
3-momentum ${\bf k_2},$ relative to the scattered--electron energy
$(y_1=\tilde\omega/\varepsilon_2)$ and $$\widetilde
L=\ln{\frac{\varepsilon^2\theta_0^{'2}}{m^2}} +2\ln{Y}\ , \ \
x_s=\frac{xyz(1+y_1)}{z-(1-y)(1+y_1)}\ ,$$
$$y_s=\frac{z-(1-y)(1+y_1)}{z}\ , \ \ Q_s^2 = Q^2z(1+y_1)\ . $$
The upper limit of the integration in Eq.~(25) $y_{1max}$ can be
defined from the condition of the inelastic--process availability
$p_x^2=(M+\mu)^2$, where $\mu$ is the pion mass. Taking into
account that for kinematics $ii)$ $q=zk_1-(1+y_1)k_2$ we obtain
$$y_{1max}=\frac{2z\varepsilon[M-\varepsilon_2(1-c)]-2M\varepsilon_2-\mu^2-2M\mu}
{2\varepsilon_2[M+z\varepsilon(1-c)]}$$ for the proton target at
rest and $$y_{1max}=\frac{2z-Y(1+c)}{Y(1+c)}$$ for the $HERA$
collider, where $c=\cos\theta.$ When writing this limit for $HERA$
we neglect the electron energy and the proton mass as compared
with the proton beam energy. Note that parameter $\theta'_0$ for
the exclusive event selection is pure auxiliary and escapes the
final result when the contribution of the region $iii)$ will be
added.

From the experimental point of view more realistic is the
calorimeter event selection, when the photon and the electron
cannot be distinguished inside narrow cone with the opening angle
$2\theta'_0$ along the outgoing--electron momentum direction.
Therefore, only the sum of the photon and electron energies can be
measured if the photon belongs to this cone. In this case we
obtain $$\frac{d\sigma^{ii),cal}_{\parallel,\bot}}{\hat y d\hat
yd\hat xdz} =
\frac{\alpha^2}{4\pi^2}P(z,L_0)\int\limits_{\Delta/Y}^{\infty}\frac{dy_1}
{(1+y_1)^3}\Bigl[\frac{1+(1+y_1)^2}{y_1}(\widetilde
L-1)+y_1\Bigr]\Sigma_{\parallel,\bot}(\hat x,\hat y,\widehat Q^2)
= $$
\begin{equation}\label{26}
\frac{\alpha^2}{4\pi^2}P(z,L_0)\Bigl[(\widetilde{L}-1)\Bigl(2\ln\frac{Y}
{\Delta}-\frac{3}{2}\Bigr)+\frac{1}{2}\Bigr]\Sigma_{\parallel,\bot}(\hat
x,\hat y,\widehat Q^2)\ .  \end{equation} For the calorimeter event
selection parameter $\theta'_0$ is the physical one, and the final result
depends on it (see below).

To calculate the contribution of the region $iii)$ we can use the
quasi--real electron method \cite{BFK} and write the leptonic
tensor in this region (that describes the radiation of collinear
photon with the energy fraction $1-z$ and noncollinear photon with
4--momentum $\tilde k$) in the following form
\begin{equation}\label{27,semicollinear tensor}
L_{\mu\nu}(k_1,k_2,(1-z)k_1,\tilde k) =
\frac{\alpha}{2\pi}P(z,L_0)\frac{dz}{z}L_{\mu\nu}(zk_1,k_2,\tilde k)\ ,
\end{equation}
$$L_{\mu\nu}(zk_1,k_2,\tilde k)= \frac{\alpha}{4\pi^2}\frac{d^3\tilde k}
{\tilde{\omega}}L_{\mu\nu}^{^{\gamma}}(zk_1,k_2,\tilde k)\ , \
L_{\mu\nu}^{^{\gamma}}(zk_1,k_2,\tilde k)=
2i\varepsilon_{\mu\nu\lambda\rho}\tilde q_{\lambda}\Bigl[
\frac{(\tilde u +\tilde t)z}{\tilde s\tilde t}k_{1\rho} + \frac{\tilde s
+\tilde u}{\tilde s\tilde t}k_{2\rho}\Bigr] \ , $$
$$\tilde q = zk_1-k_2-\tilde k\ , \ \tilde u =-2zk_2k_1\ , \
\tilde s = 2\tilde kk_2\ , \ \tilde t = -2z\tilde kk_1\ . $$

The contraction of the leptonic tensor $L_{\mu\nu}^{^{\gamma}}(k_1,k_2,
k)$ and the hadronic one, in the general case of noncollinear photon
radiation with 4-momentum $k,$ reads
\begin{equation}\label{28, || noncollinear contraction}
L_{\mu\nu}^{^{\gamma}}(k_1,k_2,k)H_{\mu\nu}^{\parallel}=
-\frac{1}{st}\bigl\{(2\tau A_t+q^2B)g_1
+2\tau[A_t-x'(u+t)B]g_2\bigr\}\frac{x'}{q^2}\ ,
\end{equation}
\begin{equation}\label{29, Perep. contraction noncoll}
L_{\mu\nu}^{^{\gamma}}(k_1,k_2,k)H_{\mu\nu}^{\bot}=-
\frac{1}{st}\Bigl\{\Bigl[A_s-\frac{uq^2}{V}B
-A_t\bigl(1-y+\frac{2u\tau}{V}\bigr)\Bigr]g_1 + \Bigl[A_s+x'(s+u)B
+
\end{equation}
$$+\bigl(1-y+\frac{2u\tau}{V}\bigr)(x'(u+t)B-A_t)\Bigr]g_2\Bigr\}
\frac{x'}{q^2}\sqrt{\frac{M^2}{Q^2}\bigl(1-y+\frac{u\tau}{V}\bigr)^{-1}}
\ , $$
$$ A_t=(u+t)^3+(uq^2-st)(u+s)\ , \ \ B=(u+t)\bigl(2V+\frac{u+t}{x'}\bigr)+
(u+s)\bigl(2V(1-y)-\frac{u+s}{x'}\bigr)\ , $$
$$A_s=(u+s)^3+(uq^2-st)(u+t)\ , \ \ q=k_1-k_2-k\ , \ \ x'
=\frac{-q^2}{2p_1q}\ , \ g_{1,2} = g_{1,2}(x',q^2)\ . $$

The contraction of the $shifted$ leptonic tensor $L_{\mu\nu}^{^{\gamma}}
(zk_1,k_2,\tilde k),$ that enters in the definition of the leptonic tensor
in the region $iii),$ and hadronic one can be obtained from Eqs.~(28) and
(29) by the substitution \begin{equation}\label{30, substitution} (k_1\ ,
k)\rightarrow (zk_1\ , \tilde k)\ , \ \ (s\ ,t\ ,u\ ,q\ ,x')\rightarrow
(\tilde s\ ,\tilde t\ ,\tilde u\ , \tilde q\ ,\tilde x) \ , \ \tilde x
=\frac{-\tilde q^2}{2p_1\tilde q}\ .  \end{equation}

We use the approach developed in Ref. \cite{AAKM} to extract the main
(proportional to $\ln\theta_0$ and $\ln\theta_0'$) contributions in
corresponding cross--section as well as to separate the infrared
singularities and write it in the following form
\begin{equation}\label{31,iii)}
\frac{d\sigma^{iii)}_{\parallel,\bot}}{\hat y d\hat xd\hat ydz}=
\frac{\alpha^2}{4\pi^2}\Bigl\{P(z,L_0)\Bigl[
\int\limits_{\Delta}^{x_{1max}}\frac{dx_1[z^2+(z-x_1)^2]}{x_1z(z-x_1)}
\ln\frac{2(1-c)}{\theta_0^2}\Sigma_{\parallel,\bot}(x_t,y_t,Q_t^2)+
\end{equation}
$$\int\limits_{\Delta/Y}^{y_{1max}}\frac{dy_1[1+(1+y_1)^2]}{y_1(1+y_1)}
\ln\frac{2(1-c)}{\theta_0^{'2}}\Sigma_{\parallel,\bot}(x_s,y_s,Q_s^2)
\Bigr]+ \frac{1+z^2}{1-z}L_0
Z_{\parallel, \bot}\Bigr\} \ , $$
where
$$x_t=\frac{xy(z-x_1)}{z-x_1+y-1}\ , \ \ y_t=\frac{z-x_1+y-1}{z-x_1}\ , \
\ Q_t^2=Q^2(z-x_1)\ .$$
For the proton target at rest
$$x_{1max} = \frac{2z\varepsilon[M-\varepsilon_2(1-c)]-2M\varepsilon_2
-\mu^2-2\mu M}{2\varepsilon[M-\varepsilon_2(1-c)]}$$
and for the $HERA$ collider conditions
$$x_{1max} = z-\frac{Y(1+c)}{2}\ .$$

The dependence on the infrared auxiliary parameter $\Delta$ as
well as on the angles $\theta_0$ and $\theta'_0$ is contained in
the first two terms on the right side of Eq.~(31), whereas the
quantities $Z_{\parallel,\bot}$ do not contain the infrared and
collinear singularities. They can be written as
\begin{equation}\label{32,Z parallel}
Z_{\parallel,\bot}=-\frac{2(1-c)}{zQ^2}
\int\limits_0^{\infty}\frac{du}{1+u^2}\Bigl\{
\int\limits_0^1\frac{dt_1}{t_1|t_1-a|}\Bigl[\int\limits_0^{x_m}
\frac{dx_1}{x_1}\Phi_{\parallel,\bot}(t_1,t_2(t_1,u))-\int\limits_0^{Yy_{1m}}\frac{dx_1}{x_1}
\Phi_{\parallel,\bot}(a,0)\Bigr] +
\end{equation}
$$\int\limits_0^a\frac{dt_1}{t_1a}\Bigl[\int\limits_0^{Yy_{1m}}
\frac{dx_1}{x_1}\Phi_{\parallel,\bot}(a,0)-\int\limits_0^{x_{1m}}\frac{dx_1}{x_1}
\Phi_{\parallel,\bot}(0,a)\Bigr]\Bigr\} \ ,$$ where we use the
same notation as in Ref. \cite{AAKM}, namely
$$t_{2,1}=\frac{1-c_{1,2}}{2}\ , \ a=\frac{1-c}{2}\ , \
t_2(t_1,u)= \frac{(a-t_1)^2(1+u^2)}{x_++u^2x_-}\ , \ c_{1,2}
=\cos{\theta_{1,2}}\ , \theta_{1,2} = {\bf{\widehat{\tilde k
k}_{1,2}}}\ ,$$ $$ x_{\pm}=t_1(1-2a)+a\pm2\sqrt{a(1-a)t_1(1-t_1)}\
.$$ Quantity $\Phi_{\parallel,\bot}(t_1,t_2)$ reads
\begin{equation}\label{33}
\Phi_{\parallel,\bot}(t_1,t_2)=\frac{\alpha^2(\tilde q^2)\tilde
x}{\widetilde Q^6} G_{\parallel,\bot} \ ,
\end{equation}
$$G_{\parallel}=g_1(2\hat\tau\widetilde A_t+\tilde q^2\widetilde
B)+ 2g_2\hat\tau(\widetilde A_t-\tilde x(\tilde u+\tilde
t)\widetilde B)\ , \ g_{1,2}=g_{1,2}(\tilde x, \tilde q^2) \ ,$$
$$G_{\bot}=\sqrt{\frac{M^2}{\widehat Q^2}(1-\hat y -\hat x\hat
y\hat\tau)^{-1}}\Bigl\{g_1\Bigl[\widetilde A_s-\frac{u\tilde
q^2}{V}\widetilde B-\widetilde A_t\bigl(1-\hat y + \frac{2u\hat
\tau}{V}\bigr)\Bigr] + $$ $$g_2\Bigl[\widetilde A_s+\tilde
x(\tilde s+\tilde u)\widetilde B+\bigl( 1-\hat y
+\frac{2u\hat\tau}{V}\bigr)[\tilde x(\tilde u+\tilde t)\widetilde
B-\widetilde A_t]\Bigr]\Bigr\} \ .$$

The upper limit of the integration $x_m$ on the right side of
Eq.~(32) is
$$\frac{2Mz\varepsilon-2M\varepsilon_2-2z\varepsilon\varepsilon_2(1-c)
-\mu^2-2M\mu}{2\varepsilon[M+z\varepsilon(1-c_1)-\varepsilon_2(1-c_2)]}\
; \ \ \frac{2z-Y(1+c)}{1+c_1}\ $$ for the proton target at rest
and the $HERA$ collider, respectively.

 \section{Total radiative correction}

\hspace{0.7cm}

The total RC to the Born cross--section (15) is defined by the sum
of the virtual and soft corrections and hard--photon emission
contribution. The last one is different for the exclusive and
calorimeter event selection. In the considered approximation it is
convenient to write this RC in the following form
\begin{equation}\label{34}
\frac{d\sigma_{\parallel,\bot}^{^{RC}}}{\hat y d\hat xd\hat
ydz}=\frac{\alpha^2}{4\pi^2}\bigl(\Sigma_{i\parallel,\bot} +
\Sigma_{f\parallel,\bot}\bigr)\ .  \end{equation} The first term
$\Sigma_i$ is independent on the experimental selection rules for
the scattered electron and reads \begin{equation}\label{35,
initial--state radiation} \Sigma_{i\parallel,\bot} =
L_0\Bigl\{\frac{1}{2}L_0P^{(2)}_{\theta}(z)+\frac{1+z^2}{1-z}\Bigl[5\ln
z-2F(z)+\ln^2Y-2\ln z\ln
Y-\frac{\pi^2}{3}+2Li_2\Bigl(\frac{1+c}{2}\Bigr)\Bigr]+
\end{equation}
$$\frac{3+z^2}{2(1-z)}\ln^2z-\frac{2(3-2z+3z^2)}{1-z}\ln(1-z)+
\frac{3-20z+z^2}{2(1-z)}\Bigr\}\Sigma_{\parallel,\bot}(\hat x,\hat
y, \widehat Q^2)+ $$
$$P(z,L_0)\ln\frac{2(1-c)}{\theta_0^2}\int\limits_0^{u_0}\frac{du}{1-u}
P^{(1)}(1-u)\Sigma_{\parallel,\bot}(x_t,y_t,Q_t^2) +
\frac{1+z^2}{1-z}L_0 Z_{\parallel,\bot}\ , \ \
u_0=\frac{x_{1max}}{z} \ , $$ where the quantity $P^{(1)}(x)$ is
defined by relations (24) and quantities $x_t,y_t,Q_t^2$ depend on
$u=x_1/z$.

The second term on the right side of Eq.~(34), denoted by
$\Sigma_f$, however, explicitly depends on the rule for the event
selection. It includes the main effect of the scattered--electron
radiation. In the case of exclusive event selection, when only the
scattered bare electron is measured, and any photon, collinear
respect to its momentum direction, is ignored, this contribution
is
\begin{equation}\label{36,excl}
\Sigma^{excl}_{f\parallel,\bot}=
P(z,L_0)\int\limits_0^{y_{1max}}dy_1[(L_Q+\ln
Y-1)P^{(1)}\bigl(\frac{1}{1+y_1}\bigr)+\frac{y_1}{1+y_1}]\Sigma_{\parallel,
\bot}(x_s,y_s,Q_s^2)\ .  \end{equation}

As it was mentioned above, in this case the parameter $\theta'_0,$
that separate kinematic regions $ii)$ and $iii),$ is not physical,
and we see that the final result does not contain it. But the mass
singularity, that is connected with the scattered--electron
radiation, exhibits itself through $L_Q$ on the right side of
Eq.~(36).

The situation is quite different for the calorimeter event
selection, when the detector cannot distinguish between the events
with a bare electron and events where the scattered electron is
accompanied by a hard photon emitted within a narrow cone with the
opening angle $2\theta'_0$ around the scattered--electron momentum
direction. For such experimental set--up we derive
\begin{equation}\label{37,cal}
\Sigma^{cal}_{f\parallel,\bot}= P(z,L_0)
\Bigl[\ln\frac{2(1-c)}{\theta_0^{'2}}\int\limits_0^{y_{1max}}
dy_1P^{(1)}\bigl(\frac{1}{1+y_1}\bigr)\Sigma_{\parallel,\bot}(x_s,y_s,Q_s^2)
+\frac{1}{2}\Sigma_{\parallel,\bot}(\hat x,\hat y,\widehat Q^2)\Bigr]\ .
\end{equation}

For the calorimeter set--up the parameter $\theta'_0$ defines the
rule of the event selection and has, therefore, the physical
sense. The final result depends on it. However, the mass
singularity due to photon emission by the final electron is
cancelled in accordance with the Kinoshita--Lee--Nauenberg theorem
\cite{KLN}. The absence of the mass singularity  indicates clearly
that term, containing $\ln\theta'_0$ on the right side of
Eq.~(37), arises due to contribution of the kinematical region
$iii)$ where the scattered electron and the photon, radiated from
the final--state, are well separated. That is why no question
appears to determine quantity $\varepsilon_2$ that enters in the
expression for $y_{1max}$.

The comparison of our analytical results for RC due to the
real--and virtual--photon emission with the analogous calculations
for unpolarized case \cite{AAKM} shows that, within the
leading-log accuracy (double--logarithmic terms in our case), this
RC are the same for the spin--dependent and spin--independent
parts of the cross--section of the radiative DIS process (1). The
difference appears on the level of the next--to--leading--log
accuracy (single--logarithmic terms in our case). That is true for
the photonic corrections in arbitrary order of the perturbation
theory.

Note that the correction to the usually measured asymmetry, which
is the ratio of the spin--dependent part of the cross--section to the
spin--independent one, is not large because the main factorized
contribution due to the virtual-- and soft--photon emission trends
to cancellation in this case. If experimental information about
the spin observables is extracted directly from the
spin--dependent part of the cross--section (for corresponding
experimental method see Ref.~\cite{G}) such cancellation does not
take place and factorized correction gives the basic contribution.

\section{The case of quasi--elastic scattering}

\hspace{0.7cm}

In previous Sections we considered the tagged--photon events in the
DIS process.  Such events can be used to measure the
spin--dependent proton structure functions $g_1$ and $g_2$ in a
single run without lowering the electron beam energy. In the
quasi--elastic limiting case, when the target proton is scattered
elastically \begin{equation}\label{38,qusielastic radiative
process} e^-(k_1)+p(p_1)\rightarrow e^-(k_2)+\gamma(k)+p(p_2) \ ,
\end{equation}
the tagged--photon events can be used also to measure the proton
electromagnetic form factors $G_E$ and $G_M$. Our final results,
obtained in Section 4, can be applied by using connection between
the spin--dependent proton structure functions $g_1$ and $g_2$ and
the proton electromagnetic form factors in this limit as given by
relations (11). Therefore, in this case we can use all formulae of
Section 4 with substitution $\Sigma^{el}_{\parallel,\bot}$ and
$G^{el}_{\parallel,\bot}$ instead of $\Sigma_{\parallel,\bot}$ and
$G_{\parallel,\bot}$ (that enters in definition of
$Z_{\parallel,\bot}$), respectively
\begin{equation}\label{39} \Sigma^{^{el}}_{\parallel}(x,y,Q^2) =
\frac{4\pi\alpha^2(Q^2)}{y(4M^2+Q^2)}
\bigl[4\tau\bigl(\tau+1-\frac{1}{y}\bigr)G_MG_E-\bigl(1-\frac{y}{2}\bigr)
(1+2\tau)G_M^2\bigr]\delta(1-x)\ ,
\end{equation}
\begin{equation}\label{40}
\Sigma^{^{el}}_{\bot}(x,y,Q^2)=\frac{8\pi\alpha^2(Q^2)}{y(4M^2+Q^2)}
\sqrt{\frac{M^2}{Q^2}[1-y(1+\tau)]}\bigl[\bigl(1-\frac{y}{2}\bigr)G_M^2
-(1+2\tau)G_MG_E\bigr]\delta(1-x) \ ,
\end{equation}
\begin{equation}\label{41}
G^{^{el}}_{\parallel,\bot} = \frac{\widetilde Q^2}{4M^2+\widetilde Q^2}
\bigl[D_{\parallel,\bot}G_M^2+E_{\parallel,\bot}G_MG_E\bigr]\delta
(1-\tilde x)\ ,
\end{equation}
$$D_{\parallel} = \bar{B}[\tilde q^2+2\hat\tau(\tilde u+\tilde
t)]\ , \
E_{\parallel}=2\hat\tau\bigl[\bigl(1+\frac{4M^2}{\widetilde
Q^2}\bigr) \widetilde A_t - \bar{B}(2\hat V+\tilde u+\tilde
t)\bigr]\ , $$ $$D_{\bot} = -K\bar B\Bigl[\frac{u\tilde
q^2}{V}+\tilde s+\tilde u+(\tilde u+ \tilde t)\bigl(1-\hat
y+\frac{2u\hat\tau}{V}\bigr)\Bigr]\ , $$
$$E_{\bot}=K\big\{\bigl(1+\frac{4M^2}{\widetilde
Q^2}\bigr)\bigl[\widetilde A_s-\bigl(1-\hat
y+\frac{2u\hat\tau}{V}\bigr)\widetilde A_t\bigr]+\bar B
\bigl[\tilde s+\tilde u(1+4\hat\tau)+(\tilde u+\tilde
t)\bigl(1-\hat y+ \frac{2u\hat\tau}{V}\bigr)\bigr]\big\}\ , $$
where $$\bar B = (\tilde u+\tilde t)(2\hat V+\tilde u+\tilde
t)+(\tilde u+\tilde s) (2\hat V(1-\hat y)-\tilde u-\tilde s)\ , \
K= \sqrt{\frac{M^2}{\widehat Q^2}(1-\hat y-\hat x\hat
y\hat\tau)^{-1}} $$ and form factors on the right side of Eq.~(41)
depend on $\tilde q^2.$

The description of the form factors is very important test for
every theoretical model of the strong interaction \cite{FC}. The
magnetic form factor of the proton $G_M$ is known with a good
accuracy in the wide interval of the momentum transfer, while the
data about the electric one $G_E$ are very poor. The recent
experiment in the Jefferson Lab on the measurement of the ratio of
the recoil--proton polarizations, performed by the Hall A
Collaboration \cite{CEBAF}, improves the situation in the region
up to $Q^2\simeq 3.5 GeV^2,$ but the region of the higher momentum
transfer remains unstudied. The use of the radiative events (38),
with both polarized and unpolarized proton target, on accelerators
with the high--intensity electron beam (for example, $CEBAF$) can
open the new possibilities in the measurement of $G_E$ as compared
with both the Rosenbluth method \cite{R} and method based on the
measurement of the recoil--proton polarizations ratio \cite{AR}.

\end{document}